\documentclass[twoside,a4paper,11pt]{sea10}
\usepackage{graphicx}
\usepackage{hyperref}
\usepackage{movie15}
\usepackage{deluxetable}

\begin{document}
\pagenumbering{arabic}
\pagestyle{myheadings}
\thispagestyle{empty}
{\flushleft\includegraphics[width=\textwidth,bb=58 650 590 680]{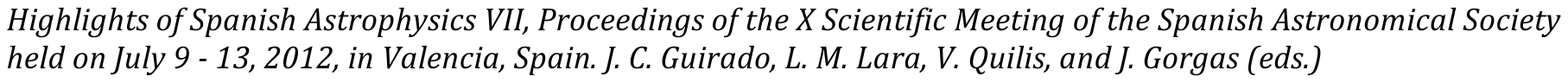}}
\vspace*{0.2cm}
\begin{flushleft}
{\bf {\LARGE
%
The ANTARES neutrino telescope
%
}\\
\vspace*{1cm}
%
Juan de Dios Zornoza$^{1}$
and Juan Z\'{u}\~{n}iga$^{1}$
%
}\\
on behalf of the ANTARES collaboration\\
\vspace*{0.5cm}
%
$^{1}$
IFIC, Instituto de F\'{i}sica Corpuscular (CSIC-Universitat de Val\`{e}ncia), Ed. Institutos de
Investigaci\'{o}n de Paterna, AC22085, 46071 Valencia, Spain\\

%
\end{flushleft}
%
\markboth{
The ANTARES neutrino telescope
}{ 
%
Juan de Dios Zornoza \& Juan Z\'{u}\~{n}iga
%
}
\thispagestyle{empty}
\vspace*{0.4cm}
\begin{minipage}[l]{0.09\textwidth}
\ 
\end{minipage}
\begin{minipage}[r]{0.9\textwidth}
\vspace{1cm}
\section*{Abstract}{\small
%

The ANTARES collaboration completed the installation of the first
neutrino detector in the sea in 2008. It consists of a three
dimensional array of 885 photomultipliers to gather the Cherenkov photons
induced by relativistic muons produced in charged-current interactions
of high energy neutrinos close to/in the detector. The scientific
scope of neutrino telescopes is very broad: the origin of cosmic
rays, the origin of the TeV photons observed in many astrophysical sources or the
nature of dark matter. The data collected up to now have allowed us to
produce a rich output of physics results, including the map of the neutrino sky of the
Southern hemisphere, search for correlations with GRBs, flaring
sources, gravitational waves, limits on the flux produced by dark
matter self-annihilations, etc. In this paper a review of these results is presented.

%
\normalsize}
\end{minipage}
%
%
%
\section{Introduction \label{intro}}

Neutrino astronomy is a powerful tool both for the astrophysics and
the particle physics fields. Neutrinos have specific advantages to
study the Universe compared to other more traditional probes, like
photons or cosmic rays. Photons, in particular at high energies, are
absorbed by matter and radiation, which severely limits their range in
the $>$1 TeV region. Cosmic rays (CRs) also interact with matter and
radiation and in addition to this, they are deflected by galactic and
extra-galactic magnetic fields. Neutrinos, on the other hand, are
neutral and only interact weakly, which makes them unique sources of
information of the high energy Universe. The price to pay, given that
they only interact weakly and that the expected fluxes are low, is
that large detection volumes are needed.

The scientific scope for neutrino telescopes is very wide. One of the
main goals is to understand the origin of CRs. A century has past
since V. Hess showed that they are produced outside the
Earth. Since they are deflected by magnetic fields, as
mentioned above, it is not clear which sources are producing them, in
particular for high energies. However, in the interactions of
high energy CRs mentioned above, high energy neutrinos are also
produced:

\begin{eqnarray} 
N + X  \longrightarrow \pi^{\pm} (K^{\pm} \dots ) + Y \longrightarrow & 
\lefteqn{\mu^{\pm} + \nu_\mu (\bar{\nu}_\mu)} \hspace{2.5cm} + Y \nonumber \\ 
 & \lefteqn{\downarrow} \hspace{3cm} \nonumber \\ 
 & \lefteqn{e^{\pm} + \bar{\nu}_\mu(\nu_\mu) + \nu_e(\bar{\nu}_e)} 
\hspace{3cm} \nonumber \\ 
\label{eq: production}
\end{eqnarray} 

Therefore, the observation of neutrinos, which are neutral and
therefore not deflected, can help to pinpoint the
sources producing CRs~\cite{bib:halzen2,bib:bednarek,bib:stecker}. 

Another important question that neutrino telescopes can help to solve
concerns the origin of the high energy photons observed in many
astrophysical sources. There are two basic mechanisms through which
these photons can be produced. One is based on the so-called leptonic
scenarios, in which the photons are produced via processes like
inverse Compton scattering. In some of the sources, this mechanism can
explain well the observations. However, this is not the case in many
sources, like supernova remnant RX1713.7 3946~\cite{bib:berezhko}. In these cases, hadronic
mechanisms can be invoked in which the observed high energy gammas are
due to the decay of neutral pions produced in the interactions of nucleons
with matter or radiation. These neutral pions will decay and produce high
energy photons. Therefore, the observation of neutrinos from the decay
of charged pions also produced in these interactions would support the hadronic scenarios.

The structure of the paper is as follows. The detection principle and
the ANTARES neutrino detector will be described in
Section~\ref{sec:antares}. Section~\ref{sec:ps} presents the results
of the search for steady point sources. The studies made on transient
sources are explained in Section~\ref{sec:transients}. The search for
diffuse cosmic fluxes are described in Section~\ref{sec:diffuse}. The
analysis looking for dark matter is presented in
Section~\ref{sec:dm}. Section~\ref{sec:other} summarizes other
searches performed by the Collaboration. Finally, the conclusions are
presented in Section~\ref{sec:conclusions}.

\section{The ANTARES telescope}
\label{sec:antares}

The detection principle of neutrino telescopes is as follows. The detector is a
three-dimensional array of photomultiplier tubes (PMTs) which collect
the Cherenkov light induced by relativistic muons produced in the
charged-current interactions of high energy muon neutrinos in the
surroundings of the detector. Other signatures are also possible:
cascade events produced in the neutral current interactions of
all the neutrino flavors or in the charged-current interactions of
electron and tau neutrinos.

\begin{figure}
\center
\includegraphics[scale=0.5]{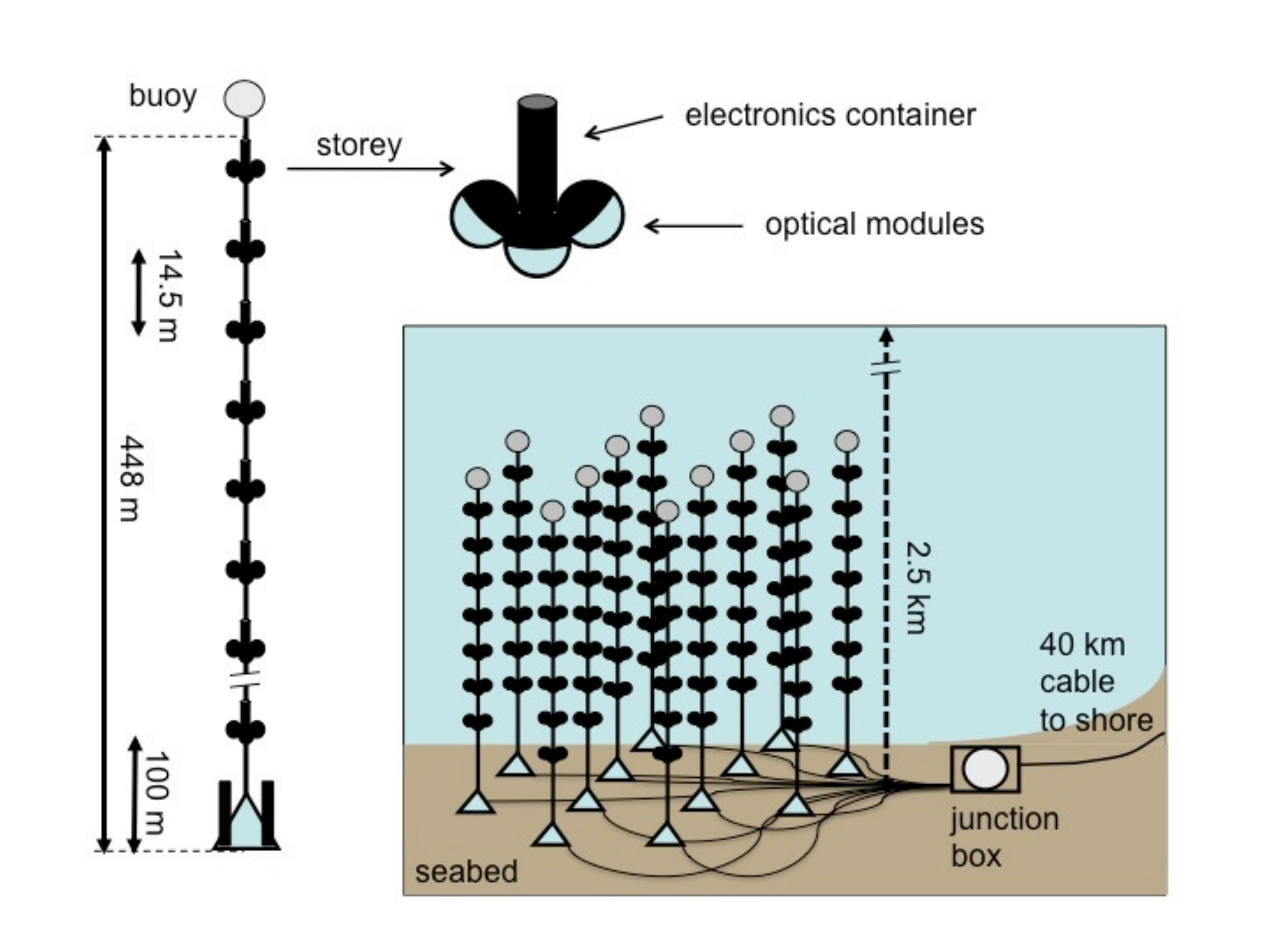} 
\caption{Schematic view of the ANTARES neutrino
  telescope. It contains 885 PMTs distributed along 12 lines anchored
  at the sea bottom.}
\label{fig:detector}
\end{figure}

ANTARES~\cite{bib:detector} is installed in the Mediterranean Sea at a
depth of 2475 m and at about 40 km from Toulon (France). It is made of
885 PMTs distributed along 12 mooring lines anchored in the bottom of
the sea. The PMTs are enclosed in glass spheres (OMs). The OMs are
grouped in triplets, being the separation between triplets (or floors)
14.5~m. The separation between lines is 60-75~m. The lines are
connected to a "junction-box" which is linked to the shore station
through an electro-optical cable. A system of hydrophones and
compasses installed in the detector keeps track of the line movements,
allowing for a resolution of about 15~cm in the position of the line
elements~\cite{bib:alignment}. In order to achieve sub-degree angular
resolution, it is also important the time
calibration~\cite{bib:timing} of the detector.  This is done by
several complementary systems, like the set of LED-based devices
called Optical Beacons. A schematic view of the detector is shown in
Figure~\ref{fig:detector}.

\section{Point source search}
\label{sec:ps}

The search for cosmic point sources is one of the main goals of
neutrino telescopes. The basic strategy for this analysis is to look
for an accumulation of events in the sky, so that the probability that
the background (atmospheric neutrinos and muons) has produced it by
chance is very low. We also use the fact that the energy spectrum for
cosmic signal is harder than for the background, so the typical energy
for signal events is higher. The cluster search is made by means of a
likelihood procedure. The background is estimated directly from data
(by randomizing the events in right ascension) in order to reduce
systematic effects. For the signal, Monte Carlo simulations provide
the point-spread function, which depends on the energy of the event. A
first analysis was done with data of 2007-2008~\cite{bib:ps1}. For the
analysis presented here, data from 29-1-2007 to 14-11-2010 are
used~\cite{bib:ps2}. The integrated live-time is 813 days, out of
which 183 correspond to the period when only five lines where
installed.

The selection of the events is based on several criteria. First, only
up-going events are selected. This condition rejects most of the
atmospheric muon background, since muons cannot traverse the
Earth. This is not enough to reject the atmospheric muon background at
an acceptable level, since a fraction of them (small, but given the
fact that the flux is very large, not negligible) are
mis-reconstructed as up-going events. However, the quality of the fit
for these misreconstructed events is bad, so a parameter describing
such reconstruction quality is used to further reject this
background. Finally, the estimated error in the reconstructed track
direction is required to be $<1^{\circ}$.  The final sample
contains 3058 neutrino candidates. Simulations predict 358$\pm$179
atmospheric muons and 2040$\pm$722 atmospheric neutrinos, consistent
with observation. The angular resolution for this sample, assuming
an E$^{-2}$ spectrum, is estimated in 0.46$\pm$0.10 degrees.

\begin{figure}
\center
\includegraphics[width=13cm,angle=0,clip=true]{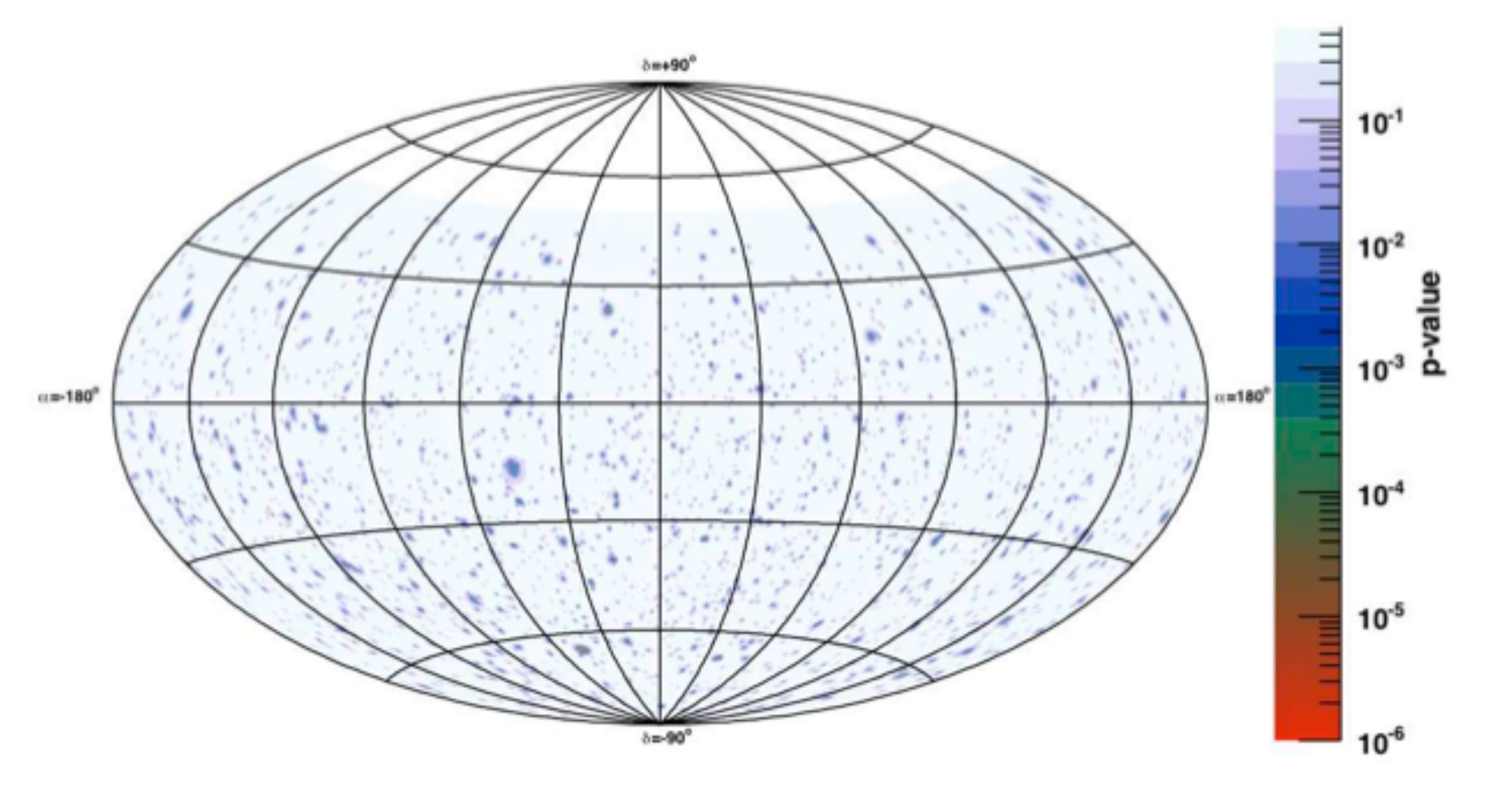} 
\caption{Skymap in equatorial coordinates of p-values obtained in the
  all-sky search. The trial factor correction is not included in the
  values shown in the figure. (Preliminary).}
\label{fig:skymap}
\end{figure}

\begin{figure}
\center
\includegraphics[width=13cm,angle=0,clip=true]{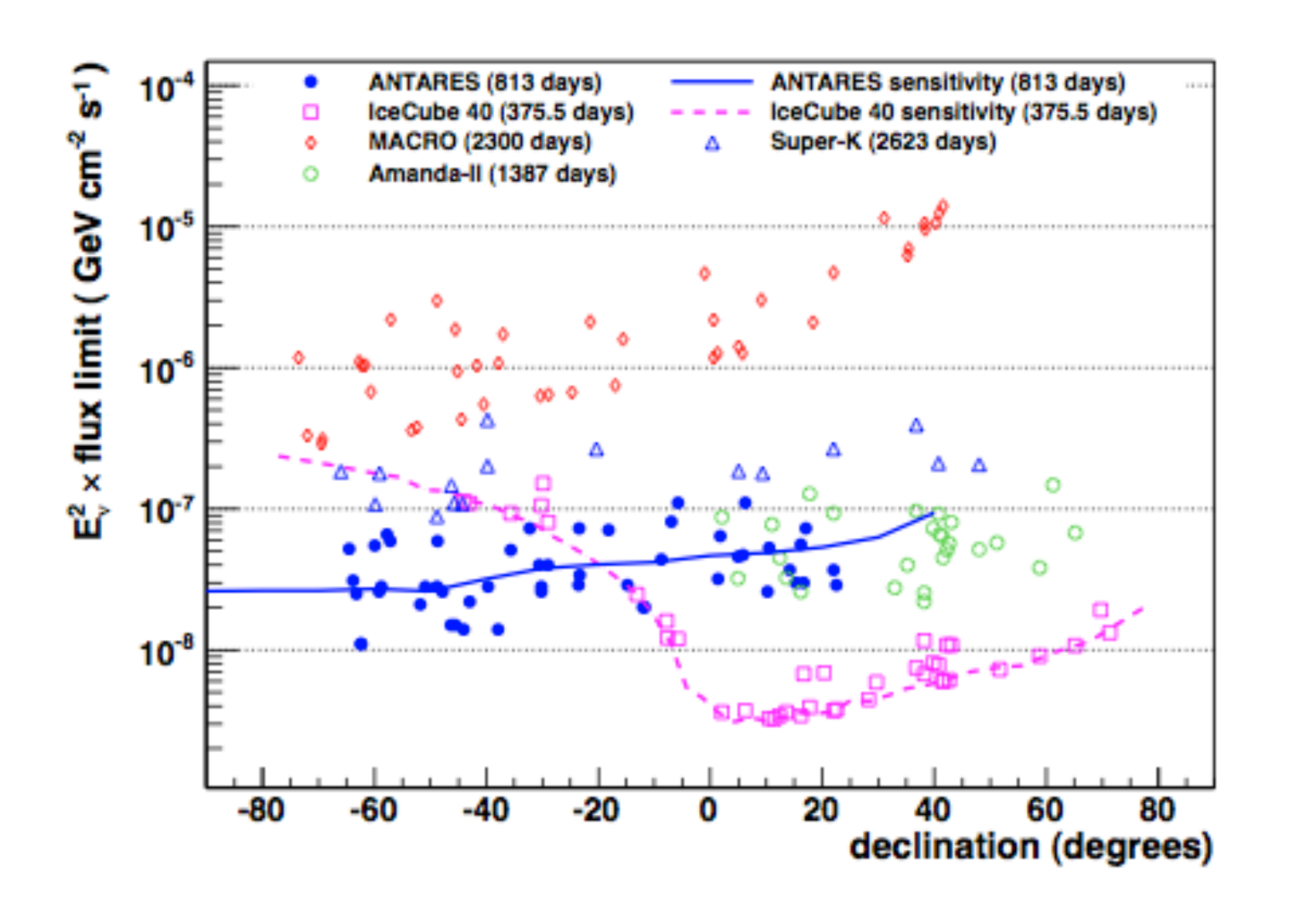} 
\caption{Upper limits in the neutrino flux (90\% c.l.) set by ANTARES
  with 2007-2010 data. The sensitivity (average upper limit) is also
  shown. Other limits set by different experiment are also shown for
  reference. An E$^{-2}$ spectrum is assumed.  (Preliminary).}
\label{fig:limits}
\end{figure}

Two different strategies have been followed for this analysis. First,
an all-sky search, where the whole sky is scanned looking for
accumulations of events. Figure~\ref{fig:skymap} shows the sky map of
the p-values before trial factor corrections. The most signal-like
cluster is found in $(\alpha, \delta)=(-46.5^{\circ}, -65.0^{\circ})$,
where there are five events within one degree around this position. By
generating pseudo-experiments for the only-background assumption, it
is estimated that the post-trial p-values is 2.6\%, equivalent to
2.2$\sigma$ significance (two-sided sigma convention). Secondly, a search in the direction of 51
sources which are good candidates for neutrino emission. No excess has
been found in any of these searches (see Table~\ref{tab:list}), being
the most significant cluster found in HESS J1023-575, with a p-value
of 0.41. Upper limits in the neutrino
flux are set, as shown in Fig.~\ref{fig:limits}.

\begin{deluxetable}{cccccccccccc}
\tabletypesize{\scriptsize}
\tablecolumns{12} 
\tablewidth{0pc} 
\tablecaption{Results from the search in the candidate list. The equatorial coordinates $(\alpha_{\rm s},\delta_{\rm s})$ in degrees, 
1$-$p-value ($\rm p$) probability and the $90\%$ C.L. upper limit on the $E^{-2}_{\nu}$ flux intensity $\phi^{90\% \rm CL}$ in units of $10^{-8} \rm GeV^{-1} cm^{-2} s^{-1}$ are
given (sorted in order of decreasing 1$-$p-value) for the 51 selected sources.}
\tablehead{ 
\colhead{Source name} & \colhead{$\alpha_s [^{\circ}]$}   & \colhead{$\delta_s [^{\circ}]$}  & \colhead{$1-\rm p$} & \colhead{$\phi^{90}CL$} 
&\colhead{   } &\colhead{   }  
&\colhead{Source name} & \colhead{$\alpha_s [^{\circ}]$}   & \colhead{$\delta_s [^{\circ}]$}  & \colhead{$1-p$} & \colhead{$\phi^{90\%CL}$}}
\startdata 
HESS J1023-575  & 155.83  & -57.76 & 0.5874970 & 6.6  & & &  SS 433           & -72.04  & 4.98   & 0.0009264 & 4.6 \\
3C 279          & -165.95 & -5.79  & 0.5231861 & 10.1 & & &  HESS J1614-518  & -116.42 & -51.82 & 0.0009264 & 2.0 \\
GX 339-4        & -104.30 & -48.79 & 0.2775692 & 5.8  & & &  RX J1713.7-3946  & -101.75 & -39.75 & 0.0009264 & 2.7  \\
Cir X-1         & -129.83 & -57.17 & 0.2135361 & 5.8  & & &  3C454.3         & -16.50  & 16.15  & 0.0009264 & 5.5 \\
MGRO J1908+06   & -73.01  & 6.27   & 0.1803912 & 10.1 & & &  W28              & -89.57  & -23.34 & 0.0009264 & 3.4  \\
ESO 139-G12     & -95.59  & -59.94 & 0.0607835 & 5.4  & & &  HESS J0632+057  &  98.24  & 5.81   & 0.0009263 & 4.6  \\
HESS J1356-645  & -151.00 & -64.50 & 0.0229395 & 5.1  & & &  PKS 2155-304     & -30.28  & -30.22 & 0.0009263 & 2.7  \\
PKS 0548-322    & 87.67   & -32.27 & 0.0146601 & 7.1  & & &  HESS J1741-302  & -94.75  & -30.20 & 0.0009262 & 2.7 \\
HESS J1837-069  & -80.59  &  -6.95 & 0.0088167 & 8.0  & & &  Centaurus\ A     & -158.64 & -43.02 & 0.0009263 & 2.1\\ 
PKS 0454-234     & 74.27   & -23.43 & 0.0015054 & 7.0 & & &  RX J0852.0-4622 & 133.00  & -46.37 & 0.0009262 & 1.5\\
IC40 hotspot & 75.45   & -18.15 & 0.0011516 & 7.0 & & & 1ES 1101-232     & 165.91  & -23.49 & 0.0009262 & 2.8 \\
PKS 1454-354     & -135.64 & -35.67 & 0.0009289 & 5.0 & & &  Vela X          & 128.75  & -45.60 & 0.0009262 & 1.5\\ 
RGB J0152+017    & 28.17   & 1.79   & 0.0009276 & 6.3 & & & W51C             & -69.25  & 14.19  & 0.0009262 & 3.6\\
Geminga          & 98.31   & 17.01  & 0.0009273 & 7.3 & & &  PKS 0426-380    & 67.17   & -37.93 & 0.0009262 & 1.4\\
PSR B1259-63     & -164.30 & -63.83 & 0.0009270 & 3.0 & & &  LS 5039          & -83.44  & -14.83 & 0.0009262 & 2.7  \\  
PKS 2005-489     & -57.63  & -48.82 & 0.0009269 & 2.8 & & &  W44             & -75.96  & 1.38   & 0.0009262 & 3.1\\
HESS J1616-508   & -116.03 & -50.97 & 0.0009268 & 2.7 & & &  RCW 86           & -139.32 & -62.48 & 0.0009262 & 1.1  \\ 
HESS J1503-582   & -133.54 & -58.74 & 0.0009268 & 2.8 & & &  Crab            & 83.63   & 22.01  & 0.0009262 & 4.1\\
HESS J1632-478  & -111.96 & -47.82 & 0.0009267 & 2.6  & & &  HESS J1507-622   & -133.28 & -62.34 & 0.0009261 & 1.1  \\ 
H 2356-309       & -0.22   & -30.63 & 0.0009266 & 3.9 & & &  1ES 0347-121    & 57.35   & -11.99 & 0.0009261 & 1.9\\
MSH 15-52       & -131.47 & -59.16 & 0.0009266 & 2.6  & & &  VER J0648+152    &  102.20 & 15.27  & 0.0009261 & 2.8  \\ 
Galactic Centre  & -93.58  & -29.01 & 0.0009266 & 3.8  & & & PKS 0537-441    & 84.71   & -44.08 & 0.0009261 & 1.3 \\
HESS J1303-631  & -164.23 & -63.20 & 0.0009265 & 2.4  & & &  HESS J1912+101   & -71.79  & 10.15  & 0.0009261 & 2.5  \\
HESS J1834-087   & -81.31  & -8.76  & 0.0009265 & 4.3  & & & PKS 0235+164    & 39.66   & 16.61  & 0.0009261 & 2.8 \\
PKS 1502+106    & -133.90 & 10.52  & 0.0009265 & 5.2  & & &  IC443            & 94.21   & 22.51  & 0.0009261 & 2.8  \\
                &         &        &           &     & & &  PKS 0727-11     & 112.58  & 11.70  & 0.0009261 & 1.9 \\

\enddata 
\label{tab:list}
\end{deluxetable} 


\section{Transient sources}
\label{sec:transients}

\begin{figure}
\center
\includegraphics[width=13cm,angle=0,clip=true]{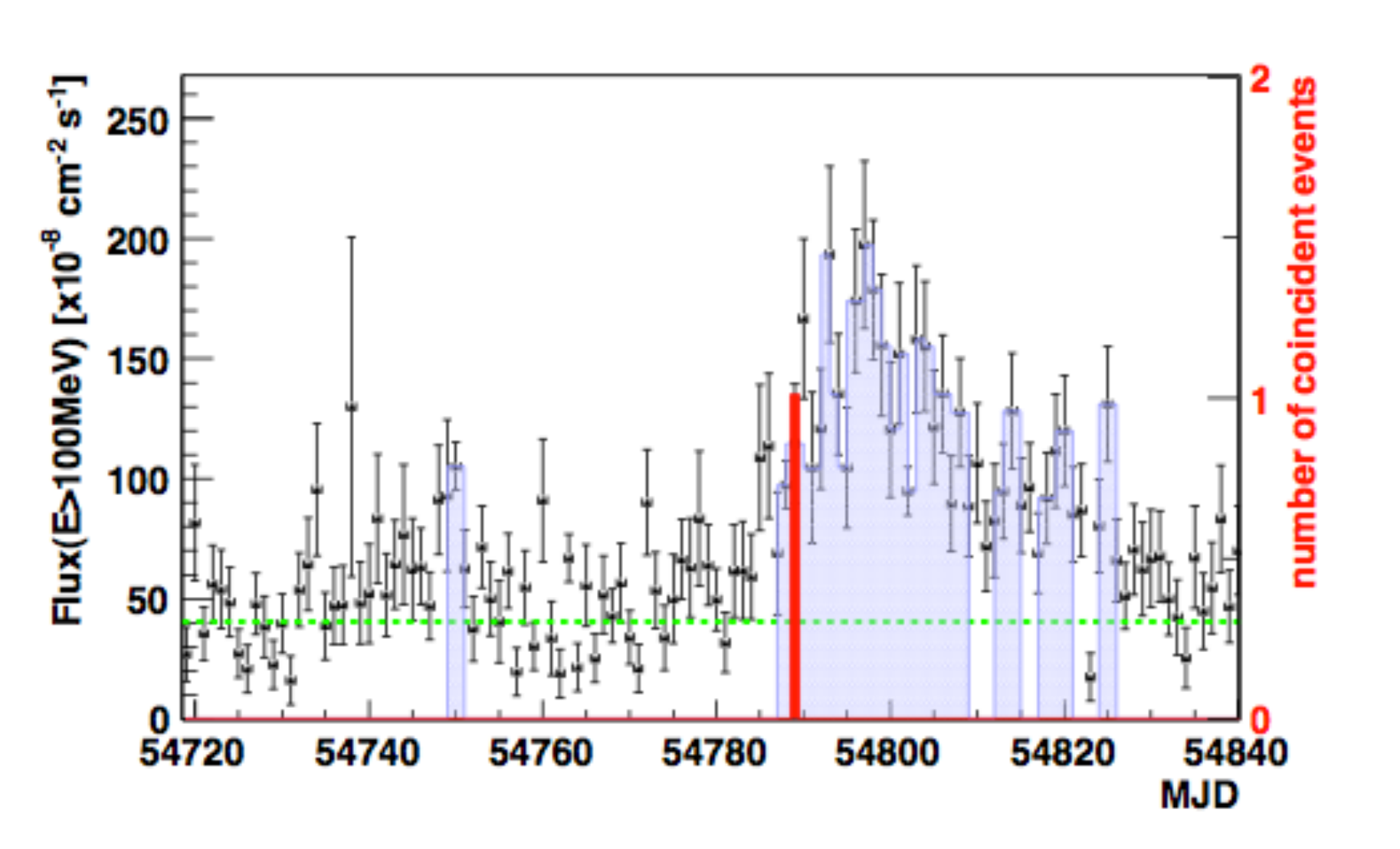} 
\caption{Gamma ray light curve of 3C279 as observed by the LAT instrument in
  the Fermi satellite. The red line indicates the time of the event
  observed by ANTARES in correlation with this flare.}
\label{fig:flare}
\end{figure}

For many astrophysical sources, the time information can be an
additional criterion for background rejection and therefore
improving the sensitivity. This is the case of catastrophic events
like GRBs or flaring objects like blazars or micro-quasars. In the GRB
analysis 40 bursts detected in 2007 have been investigated looking for
correlations, with negative results. In the case of the blazar
analysis~\cite{bib:flares}, 10 flares detected in 2008 have been analyzed. In nine of
them (PKS0208-512, AO0235+164, PKS1510-089, 3C273, 3C279, 3C454.3,
OJ287, PKS0454-234, Wcomae and PKS2155-304), no event has been found
in correlation with them. In the case of 3C279, one event is found in
correlation. The corresponding p-value, taking into account the trial
factor, is 0.1. Figure~\ref{fig:flare} shows the light curve for the
flare of 3C279 and the time at which the neutrino event is
found. Finally, in the micro-quasar analysis, 6 flares detected in
2007-2012 have been analyzed (Circinus X-1, GX339-4, H 1743-322,
IGRJ17091-3624, Cygnus X-1, Cygnus X-3), with no event
correlated it time.

\section{Diffuse fluxes}
\label{sec:diffuse}

\begin{figure}
\center
\includegraphics[width=13cm,angle=0,clip=true]{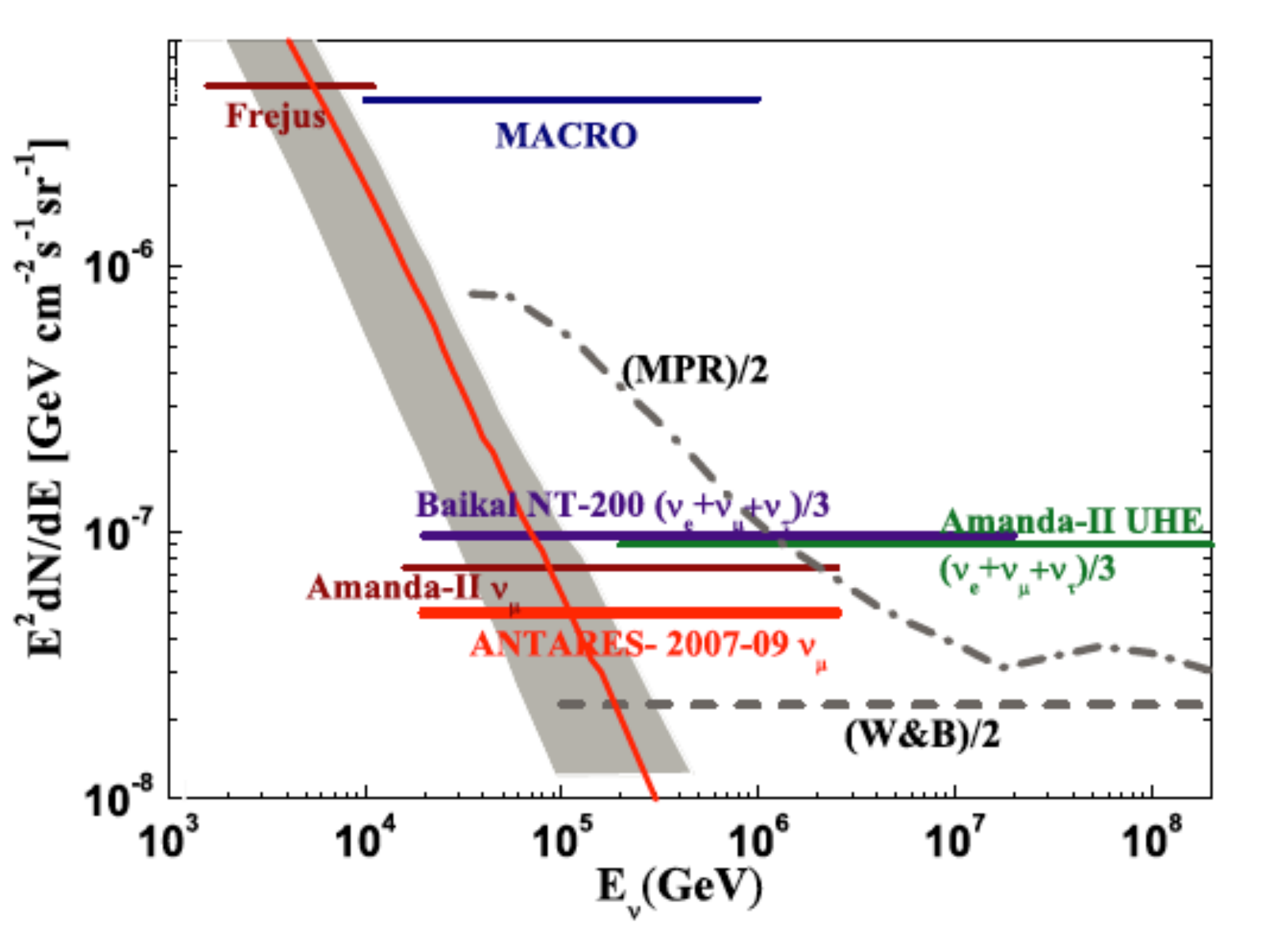} 
\caption{Upper limits at 90\% c.l. for the diffuse flux contribution
  of neutrinos and anti-neutrinos. The grey band represent the
  expected atmospheric neutrino flux. Its width shows the dependence of
the atmospheric flux with the zenith angle.}
\label{fig:diffuse}
\end{figure}

An alternative approach in the search for cosmic sources is to
integrate all the signal of the observable sky, i.e. to sum up the
contributions from all the unresolved sources. The price to pay in
this case is that the background cannot be rejected with the
directional information and therefore is larger. However, the fact
that the energy spectrum expected for cosmic sources is harder
($E^{-2}$) than for the atmospheric background ($E^{-3.7}$) allows for
a discrimination based on some energy-depending variable. For
the analysis presented here, a variable based on the number of hit
repetitions in a given optical module is
used. Figure~\ref{fig:diffuse} shows the results. Since no excess has
been observed, a limit on the cosmic diffuse flux is set~\cite{bib:diffuse}.

\section{Dark matter}
\label{sec:dm}

\begin{figure}
\center
\includegraphics[width=13cm,angle=0,clip=true]{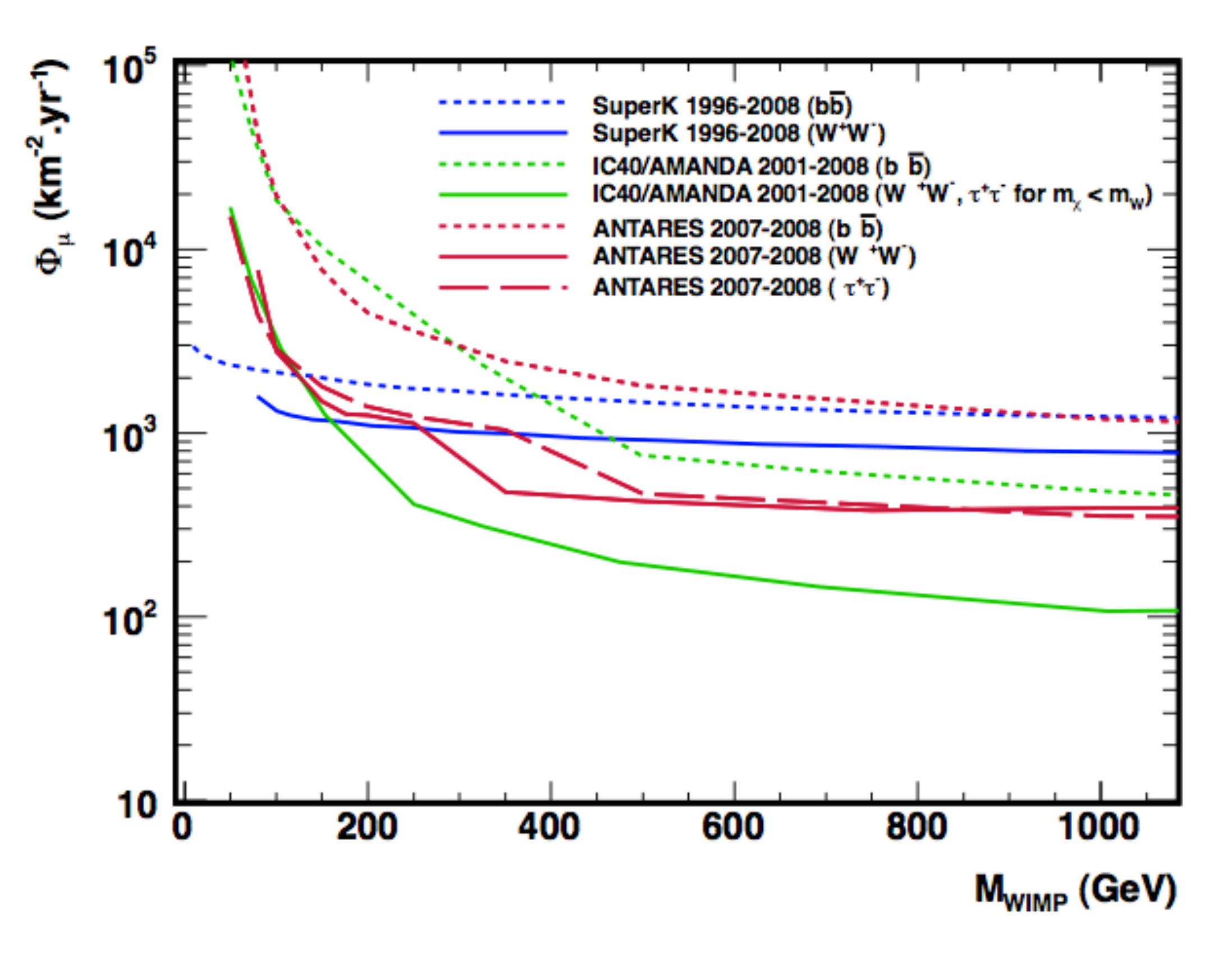} 
\caption{Limits in the neutrino flux (90\% c.l.) from the Sun for
  several channels in the CMSSM and mUED frameworks.  (Preliminary).}
\label{fig:dm}
\end{figure}

The search for dark matter is also one of the main goal of neutrino
telescopes. They have specific advantages both when compared with
other indirect searches and with direct detection experiments. If dark
matter is made of WIMPs (Weakly Interacting Massive Particles), these
particles would scatter in massive objects like the Sun or the Earth,
lose energy and become gravitationally trapped, accumulating in the
centre of the object. The Galactic Centre is also expected to have a
large density of dark matter particles. After the self-annihilations
of these WIMPS, high energy neutrinos are produced (either
indirectly, i.e. from the secondaries produced in the WIMP
self-annihilations, or directly, like in some Kaluza-Klein
scenarios). The potential detection of a high energy neutrino signal
from the Sun, for instance, would be a very clean signal of dark
matter, since no other known astrophysical explanation would be able
to explain it. This is not the case of observations of gamma-ray or
positron excess observed by other indirect searches. Moreover, though
the sensitivities for spin-independent cross sections cannot compete with
those of direct search experiments (since it is proportional to the
square of the atomic number of the target and in the Sun there are
mostly protons and helium), this is not the case for spin-dependent
cross section, where neutrino telescopes offer the best limits. The
analysis made by the ANTARES collaboration is a binned search in the
direction of the Sun, using data of 2007-2010~\cite{bib:dm}. No signal has been
observed, so limits in the flux are set, as shown in Fig.~\ref{fig:dm}.

\section{Other searches}
\label{sec:other}

Here we summarize other analyses performed by ANTARES:
\begin{itemize}
\item Gravitational waves: Several catastrophic astrophysical events
  are good candidates to emit both gravitational waves and
  neutrinos. A first analysis looking for correlations with events of
  VIRGO and LIGO has been done using 2007 data (where
  only five lines of ANTARES where installed). A second analysis,
  using data of 2009-2010 is ongoing~\cite{bib:gw}.
\item Magnetic monopoles: Magnetic monopoles, predicted in
  spontaneously broken gauge theories, would be detected in
  ANTARES as very bright ($\sim$8500 times the light output of a muon)
  and slowly moving events. Flux limits have been set in the range
  1-9$\times10^{-17}$ cm$^{-2}$ s$^{-1}$ sr$^{-1}$, a factor three
  better than previous searches by other experiments~\cite{bib:monopoles}.
\item Nuclearites: Nuclearites are stable, massive lumps of strange
  quark matter particles. The signature for identifying these events
  are slowly moving downgoing events. The data taken in 2007-2008 has
  allowed us to set flux limits at 1-5$\times10^{-17}$ cm$^{-2}$
  s$^{-1}$ sr$^{-1}$ for nuclearite masses in the range
  10$^{14}$-10$^{17}$ GeV~\cite{bib:nuclearites}.
\item Fermi bubbles: The so-called Fermi bubbles are two large almost
  spherical structures located above and below the Galactic plane,
  close to the Galactic Centre. Fermi-LAT has observed gamma emission
  with a hard an uniform spectrum. There are models which explain
  these structures as a consequence of hadronic acceleration. A on/off
  source search has found 75 events in the on-source region, whereas
  90$\pm$5(stat)$\pm$3(sys) were expected from background. This
  excludes the fully hadronic scenario with no cut-off~\cite{bib:fermi}.
\item Neutrino oscillations: Although the ANTARES detector does not aim for
  oscillations measurements and therefore its sensitivity cannot
  compete with other experiments specifically designed for such
  aim, it is interesting to note oscillations have been observed
  by ANTARES, with results compatible with the other experiments~\cite{bib:oscillations}.
\end{itemize}

\section{Conclusions}
\label{sec:conclusions}

Neutrino astronomy has become a mature field and constitutes a
powerful tool both in Astrophysics and Particle Physics. The ANTARES
collaboration has successfully completed the construction of the first
neutrino telescope in the sea. Being in the Northern Hemisphere, it can
observe most of the Southern sky, including the Galactic Centre, with
unsurpassed sensitivity. After five years of data taking, it has
produced a rich scientific output. The physics results obtained up to now
include the following searches: steady point sources, including the first map of the
Southern neutrino sky; correlations with flaring sources, like
micro-quasars and blazars; coincidences with GRBs; diffuse cosmic
fluxes; correlation with gravitational waves; dark matter from the
Sun, setting constraints in frameworks with neutrino and Kaluza-Klein
particles; nuclearites, and magnetic monopoles. Even if the results of
the searches has not been positive yet, the project has shown the
feasibility of such a kind of detectors and therefore has paved the
way to the next step, a cubic kilometer detector, KM3NeT.

%
%
\small  
%
\section*{Acknowledgments}   
%
The authors acknowledge the financial support of the funding agencies:
Ministerio de Ciencia e Innovaci\'{o}n (MICINN), Prometeo of Generalitat
Valenciana and MultiDark, Spain

%

%

\begin{thebibliography}{}
\small
%

\bibitem{bib:ps1} S. Adri\'an-Mart\'inez {\it et al.}, ANTARES
Collaboration, 
Ap. J. {\bf 743}, (2011) L14, [arXiv:1108.0292].

\bibitem{bib:flares}  S. Adri\'an-Mart\'inez {\it et al.}, ANTARES
Collaboration, 
Accepted by Astropart. Phys., [arXiv:1111.3473].

\bibitem{bib:monopoles} S. Adri\'an-Mart\'inez {\it et al.}, ANTARES Collaboration,
 Astropart. Phys. {\bf 35} (2012), 634-640, [arXiv:1110.2656].

\bibitem{bib:alignment} S. Adri\'an-Mart\'inez {\it et al.}, ANTARES Collaboration, 
JINST {\bf 7} (2012), T08002, [arXiv:astro-ph/1202.3894].

\bibitem{bib:oscillations}  S. Adri\'an-Mart\'inez {\it et al.}, ANTARES Collaboration,
Phys. Lett. {\bf B714} (2012), 224-230, [arXiv: 1207.3105].

\bibitem{bib:gw}  S. Adri\'an-Mart\'inez {\it et al.}, ANTARES, VIRGO
 and LIGO Collaboration,
[arXiv:1205.3018].

\bibitem{bib:ps2} S. Adri\'an-Mart\'inez {\it et al.}, ANTARES Collaboration,
Submitted to Ap.J., [arXiv:1207.3105].

\bibitem{bib:timing} J. A. Aguilar {\it et al.}, ANTARES
Collaboration, 
Astropart. Phys. {\bf 34} (2011) 539-549,
[arXiv:1012.2204].

\bibitem{bib:diffuse} J. A. Aguilar  {\it et al.}, ANTARES Collaboration,
Phys. Lett. {\bf B696}, (2011), 16-22, [arXiv:1011.3772].

\bibitem{bib:detector} J. A. Aguilar {\it et al.}, ANTARES Collaboration,
Nuclear Instrum. Meth. in Physics Research, {\bf A 656} (2011), 11-38,
[arXiv:1104.1607v1].

\bibitem{bib:bednarek} W. Bednarek {\it et al.},
New Astron. Rev. {\bf 49} (2005) 1, [arXiv:astro-ph/0404534].

\bibitem{bib:berezhko} E. G. Berezhko, H. J. V\"olk, 
Astronomy and Astrophysics, {\bf 492}, 3 (2008), 695-701, [arXiv:0810.0988v2].

\bibitem{bib:halzen2} F. Halzen, D. Hooper,
Rep. Prog. Phys. {\bf 65} (2002) 1025,  [arXiv:astro-ph/0204527].

\bibitem{bib:fermi} V. Kulikovskiy,
Neutrino 2012 Conference, Kyoto 2012.

\bibitem{bib:nuclearites} G. Pavalas,
Proc. of the CSSP10 conference, Sinaia (2010), [arXiv:1010.2071v1].

\bibitem{bib:stecker} F. W. Stecker, 
 Phys. Rev. D {\bf 72} (2005) 107301, [arXiv: astro-ph/0510537]

\bibitem{bib:dm} J.D. Zornoza,
Nucl. Instrum. Meth. {\bf A 692} (2012) 123-126, [arXiv:1204.5066].









%
%
\end{thebibliography}
\end{document}